\documentclass[prd,twocolumn,superscriptaddress,showpacs,amsmath,amssymb,nofootinbib]{revtex4-1}
\pdfoutput=1
\usepackage{graphicx,epsf,amsmath}
\usepackage[]{latexsym}
\usepackage{color}
\newcommand{\be}{\begin{eqnarray}}
\newcommand{\ee}{\end{eqnarray}}

\newcommand{\pro}[2]{\mbox{$\langle\, #1 \mid #2\,\rangle$}}
\newcommand{\expec}[1]{\mbox{$\langle\, #1\,\rangle$}}

\renewcommand{\d}{\mbox{${\rm d}$}}
\newcommand{\lp}{\ell_{\rm p}}
\newcommand{\mpl}{m_{\rm p}}

\newcommand{\rh}{R_{\rm H}}
\begin{document}
%
%
\title{Horizon Wave-Function and the Quantum Cosmic Censorship}
\author{Roberto~Casadio}
\email{casadio@bo.infn.it}
\affiliation{Dipartimento di Fisica e Astronomia,
Alma Mater Universit\`a di Bologna,
via~Irnerio~46, 40126~Bologna, Italy}
\affiliation{I.N.F.N., Sezione di Bologna, viale Berti~Pichat~6/2, 40127~Bologna, Italy}
\author{Octavian~Micu}
\email{octavian.micu@spacescience.ro}
\affiliation{Institute of Space Science, Bucharest,
P.O.~Box MG-23, RO-077125 Bucharest-Magurele, Romania}
\author{Dejan Stojkovic}
\email{ds77@bufalo.edu}
\affiliation{HEPCOS, Department of Physics, SUNY at Buffalo, Buffalo, NY 14260-1500}
\begin{abstract}
We investigate the Cosmic Censorship Conjecture by means of
the horizon wave-function (HWF) formalism.
We consider a charged massive particle whose quantum mechanical state
is represented by a spherically symmetric Gaussian wave-function,
and restrict our attention to the superxtremal case (with charge-to-mass ratio
$\alpha>1$), which is the prototype of a naked singularity in the classical 
theory. We find that one can still obtain a normalisable HWF for $\alpha^2<{2}$,
and this configuration has a non-vanishing probability of being a black hole,
thus extending the classically allowed region for a charged black hole. 
However, the HWF is not normalisable for $\alpha^2 > 2$,
and the uncertainty in the location of the horizon blows up at $\alpha^2=2$,
signalling that such an object is no more well-defined.
This perhaps implies that a {\em quantum\/} Cosmic Censorhip might be conjectured
by stating that no black holes with charge-to-mass ratio greater than a critical
value (of the order of $\sqrt{2}$) can exist.
\end{abstract}
\maketitle
\section{Introduction}
A complete understanding of the gravitational collapse of a compact object remains one
of the most challenging issues in contemporary theoretical physics. 
The general relativistic (GR) description, resulting in the
formation of a black hole (BH) or naked singularity (NS), was first investigated in the
papers of Oppenheimer and co-workers~\cite{OS}.
Although the literature on the subject has grown immensely (see, e.g.~Ref.~\cite{joshi}),
many technical and conceptual issues remain. 
One of these is the famous Cosmic Censorship Conjecture (CCC), proposed by Penrose in
1969~\cite{Penrose:1969pc}, which states that no singularities will ever become visible
to an outer observer in a generic gravitational collapse starting from reasonable
nonsingular initial states.
To date, the conjecture remains unproved, and it is considered one of the most
important open problems in gravitational physics.
Another great open issue in GR is the problem of considering the quantum mechanical
(QM) nature of the collapsing
matter~\cite{Greenwood:2008ht}.
We will here address both issues for the Reissner-Nordstr\"om (RN) geometry,
which describes charged BHs, a subject of many theoretical investigations in the
past (see, e.g.~Ref.~\cite{Wang:2009ay}).
\par
Most attempts at quantising BH metrics consider the gravitational degrees of freedom
unrelated to the matter state that sources the geometry.  
More recently, the Horizon Wave Function (HWF) formalism was proposed~\cite{Casadio},
as a way of quantising the Einstein equation that determines the gravitational radius
of a spherically symmetric matter source and its time evolution~\cite{C14}, which instead
relates the quantum state of the horizon to the quantum state of matter. 
This formalism was then applied to a few different case studies~\cite{CGUP,Ctest,BEC_BH},
yielding apparently sensible results in agreement with (semi)classical expectations, and
there is therefore hope that it will facilitate our understanding of the formation of BHs
from QM particles.
In particular, it seems natural to extend this formalism beyond
Schwarzschild BHs and tackle the CCC from a quantum perspective
by considering an electrically charged particle represented by
a Gaussian wave-packet in the classical regime in which it would be a NS. 
\section{Electrically charged spherical sources}
We start by recalling the classical RN metric can
be written as
\be
\d s^2
=
-f\,\d t^2
+f^{-1}\,\d r^2
+r^2
\left(d\theta^2 + \sin^2 \theta\,d\phi^2\right)
\ ,
\label{RN}
\ee
with 
\be
f=
1- \frac{2\, \lp\, M}{\mpl \,r}+\frac{Q^2}{r^2}
\ ,
\ee
where $M$ and $Q$ respectively represent the ADM mass and charge of the source,
$\lp$ is the Planck length and $\mpl$ the Planck mass~\footnote{We shall use units
with $c=k_B=1$, and always display the Newton constant $G=\lp/\mpl$,
so that $\hbar=\lp\,\mpl$.}.
For $|Q|<\lp\,M/\mpl$, the above metric contains two horizons, namely 
\be
R_{\pm}
=
\lp\, \frac{M}{\mpl}\pm\sqrt{\left(\lp\, \frac{M}{\mpl}\right)^2 - Q^2}
\ ,
\label{R+-}
\ee
and represents a BH.
The two horizons overlap for $|Q|=\lp\,M/\mpl$, the so-called extremal BH case,
while for $|Q|>\lp\,M/\mpl$ no horizon exists and the
central singularity is therefore accessible to outer observers.
This is the prototype of a NS, which we will refer to as the
``superextremal geometry''.
It is in fact more convenient to express all relevant quantities
in terms of the mass $M$ and the (positive definite) specific charge
\be
\alpha = \frac{|Q|\, \mpl}{\lp\, M}
\ .
\label{alpha}
\ee
Using this parameter, the above expression~\eqref{R+-} becomes
\be
R_{\pm}
=
\lp\, \frac{M}{\mpl}\left(1\pm\sqrt{1- \alpha^2}\right)
\ ,
\label{Ralpha}
\ee
and the three regimes mentioned above are then explicitly
parametrised as i) $0<\alpha < 1$ for the BH with two horizons~\footnote{We
remark in passing that the inner horizon gives rise to an instability usually
referred to as ``mass inflation'',
but we have investigated under which conditions the inner horizon
is actually realised in the quantum context in a separate work~\cite{InProg}.},
ii) $\alpha = 1$ for the extremal BH, and iii) $\alpha > 1$ for the
superextremal geometry. 
\par
We shall now investigate the superextremal geometry from
a quantum mechanical perspective by first determining the HWF
for $\alpha<1$ and then extending it continuously into 
the regime $\alpha>1$.
\subsection{HWF for Gaussian sources}
The general procedure that leads to the HWF~\cite{Casadio,C14}
is based on lifting the gravitational radius $\rh$ of a spherically symmetric QM system
to the rank of a quantum operator.
This step can be physically motivated by first recalling that the coordinate $r$ in 
a spherical metric like the one in Eq.~\eqref{RN} is invariantly related to the geometrical
area $4\,\pi\,r^2$ of a sphere centred on the origin $r=0$, and is therefore a natural
candidate to become an observable in the quantum theory.
Moreover, the specific property that qualifies $r=\rh$ is that it represents
the location of trapping surfaces and thus determines the causal structure of the
space-time, which one can assume will also ought to remain an observable property
in the quantum theory. 
In details, we recall that in a neutral spherically symmetric system,
$\rh(r)=2\,\lp\,M(r)/\mpl$, where
\be
M(r,t)
=
4\,\pi\int_0^r \rho(\bar r,t)\,\bar r^2\,\d \bar r
\ ,
\label{M}
\ee
is the Misner-Sharp mass.
One should notice that $M$ represents the {\em total\/} energy (thus, roughly
speaking, including the negative gravitational energy) and is related to the
energy density $\rho$ of the source via the {\em flat\/} space volume.
A specific value of $r$ is then a trapping surface if $\rh(r)=r$,
whereas, if $\rh(r)<r$, the gravitational radius is still well-defined but
does not correspond to any peculiar causal surface.
In the electrically charged case, we have two gravitational radii and
corresponding operators, namely $\hat R_\pm$.
The classical relation~\eqref{Ralpha} will then be reinterpreted in this
context as the operatorial equation relating $\hat R_\pm$ to the total energy
$\hat M$ of the system, with $\hat R_\pm$ acting multiplicatively on
the HWF and $\hat M$ acting multiplicatively on energy eigenstates.
Finally, we will consider the ratio $\alpha$ as a simple parameter.
\par
Let us consider as a source of the RN space-time an electrically
charged massive particle at rest in the origin of the reference frame,
represented by a spherically symmetric Gaussian wave-function
\be
\psi_{\rm S}(r)
=
\frac{e^{-\frac{r^2}{2\,\ell^2}}}{\ell^{3/2}\,\pi^{3/4}}
\ .
\label{psis}
\ee
We emphasise that the radial variable $r$ in the above is interpreted 
to be the same as the coordinate $r$ in the metric~\eqref{RN}, and is therefore a
measure of the particle's size as seen from an outer observer who, e.g.~scatters
particles against it.
We shall also assume that the width of the Gaussian $\ell$
is the minimum compatible with the Heisenberg uncertainty principle,
that is~\footnote{A thorough analysis of the changes that occur
by relaxing this condition can be found in Ref.~\cite{C14}.}
\be
\ell
=
\lambda_m
\simeq
\lp\,\frac{\mpl}{m}
\ ,
\ee
where $\lambda_m$ is the Compton length of the particle of rest mass $m$.
In momentum space, the wave-function of the particle described above is 
\be
\psi_{\rm S}(p)
=
\frac{e^{-\frac{p^2}{2\,\Delta^2}}}{\Delta^{3/2}\,\pi^{3/4}}\, \label{psi_p}
\ ,
\label{psip}
\ee
where $p^2=\vec p\cdot\vec p$ is the square modulus of the spatial momentum,
and the width
$
\Delta
=
\mpl\,{\lp}/{\ell}
\simeq
m
$.
For the energy of the particle, we shall employ the relativistic mass-shell relation
in flat space,
\be
M^2=p^2+m^2 
\ ,
\label{mass shell}
\ee
in analogy with the expression of the Misner-Sharp mass~\eqref{M}.
This choice does not therefore imply that the effects of curved space-time are
{\em a priori\/} discarded, but rather that they should be included in the very definition
of total energy of the system~\footnote{Let us also remark that, for a source
of Planckian mass and size, quantum effects might make the very concept of ``curved space-time''
inadequate in its proximity, whereas determining (asymptotic) cross sections for particle
collisions might still be meaningful. 
\label{f4}}.
\par
For $\alpha<1$, it is clear that one can now write a HWF for each of the two horizons.
In fact, from the quantum version of Eq.~\eqref{R+-}, the total energy $M$ can be
expressed in terms of the two horizon radii as
\be
\lp\,\frac{\hat M}{\mpl}
=
\frac{\hat R_{+}+\hat R_{-}}{2}
\ ,
\label{EofRpm}
\ee
and
\be
\hat R_{\pm}
=
\hat R_{\mp}\,
\frac{1\pm\sqrt{1-\alpha^2}}{1\mp\sqrt{1-\alpha^2}}
\ .
\label{R-R+}
\ee
Note that we promoted $M$, $R_{+}$, and $R_{-}$ into operators
$\hat{M}$, $\hat R_{+}$, and $\hat R_{-}$, which are related to the
corresponding observables.
Our specific choice is not unique, and it is associated with usual
ambiguities when going from a classical to quantum formalism.  
\par
The unnormalised HWFs for $R_{+}$ and $R_{-}$ are then obtained 
by expressing $p$ from the mass-shell relation~\eqref{mass shell}
in terms of the energy $M$ in Eq.~\eqref{EofRpm},
and then replacing one of the relations in
Eq.~\eqref{R-R+} into Eq.~\eqref{psi_p}.
The two HWFs are then given by
\be
\psi_{\rm H}(R_{\pm})
&=&
\mathcal{N_\pm}\,
\Theta\left(R_{\pm}-R_{\rm min \pm}\right)
\nonumber
\\
&&
\times\, 
\exp\left\{-\frac{\mpl^2\,R_{\pm}^2}{2\,\Delta^2\,\lp^2\,(1\pm\sqrt{1-\alpha^2})^2}\right\}
\ ,
\label{psih}
\ee
where the Heaviside function arises from the minimum energy in the spectral
decomposition of the wave-function~\eqref{psis} being $M=m$, which
corresponds to 
\be
R_{\rm min \pm}
=
\lp\,\frac{m}{\mpl}\left(1\pm\sqrt{1 - \alpha^2}\right)
\ .
\label{Rminalpha}
\ee
Finally, the normalizations $\mathcal{N_\pm}$ are fixed by assuming the scalar
product~\footnote{The analytical expressions for the normalization of the HWFs
are very cumbersome and not particularly significant, thus we will omit them
throughout the paper.} 
\be
\pro{\psi_{\rm H}}{\phi_{\rm H}}
=
4\,\pi\int_{0}^\infty \!\!
\psi_{\rm H}^*(R_{\pm})\,\phi_{\rm H}(R_{\pm})\,R_{\pm}^2\,\d R_{\pm}
\ ,
\label{normH}
\ee
where, like in the previous equation, the upper signs are used for the
normalization of $\psi_{\rm H}(R_{+})$, while the lower signs are used
when normalizing $\psi_{\rm H}(R_{-})$.
\par
The probability density that the particle lies inside its horizon of radius
$r=R_{\pm}$ can now be calculated starting from the wave-functions~\eqref{psih}
associated with~\eqref{psis} as
\be
{\mathcal P}_{<\pm}(r<R_{\pm})
=
P_{\rm S}(r<R_{\pm})\,{\mathcal P}_{\rm H}(R_{\pm})
\ ,
\label{PrlessH}
\ee
where
\be
P_{\rm S}(r<R_{\pm})
=
4\,\pi\,\int_0^{R_{\pm}}
|\psi_{\rm S}(r)|^2\,r^2\,\d r
\ee
is the probability that the particle is inside a sphere of radius $r=R_{\pm}$,
and
\be
{\mathcal P}_{\rm H}(R_{\pm})
=
4\,\pi\,R_{\pm}^2\,|\psi_{\rm H}(R_{\pm})|^2
\label{Ph}
\ee
is the probability density that the sphere of radius $r=R_{\pm}$
is a horizon.
Finally, one can integrate~\eqref{PrlessH} over all possible
values of the horizon radius $R_+$ to find the probability for the particle
described by the wave-function~\eqref{psis} to be a BH, namely
\be
P_{\rm BH+}
=
\int_{R_{\rm min+}}^\infty {\mathcal P}_{<+}(r<R_+)\,\d R_+
\ .
\label{PBH+}
\ee
The analogous quantity for $R_-$,
\be
P_{\rm BH-}
=
\int_{R_{\rm min-}}^\infty {\mathcal P}_{<-}(r<R_-)\,\d R_-
\ ,
\label{PBH-}
\ee
will instead be the probability that the particle lies further inside 
its inner horizon, and both $R_-$ and $R_+$ are therefore realized
(for more details about this case, see Ref.~\cite{InProg}).
\par
Before moving to the specific topic of this work,
let us note that for a particle with mass $M\simeq m$ smaller than
the Planck scale, the classical values of $R_\pm$ from Eq.~\eqref{Ralpha}
are smaller than the Planck length.
This however should not prevent one from considering the 
corresponding $\hat R_\pm$ as valid quantum operators representing
the gravitational radii of the system.
In fact, let us first point out that, for any value of the particle's mass,
if $\expec{\hat R_\pm}\lesssim \ell$, these radii most likely do not
correspond to horizons, exactly like in the classical theory there is
no trapping surface of size $r$ if the corresponding gravitational radius
$R(r)=2\,\lp\,M(r)/\mpl<r$.
Consequently, the probability that the particle is a BH decreases
very fast below the Planck mass, as it was shown for the
simpler case of an electrically neutral particle in Refs.~\cite{Casadio,CGUP,C14},
and is confirmed for the probability $P_{\rm BH+}$ of the RN BH in
Ref.~\cite{InProg}.
This means that, although one can have $\expec{\hat R_\pm}\lesssim \lp$,
QM fluctuations due to the uncertainty in the particle's size
dominate below the Planck energy and the horizon is simply not
realised as an actual trapping surface (see also footnote~\ref{f4}).
More realistically, it was shown in Ref.~\cite{Ctest}, that a BH
has a significant probability to form from the collision of sub-Planckiian
particles only if the center-of-mass energy reaches into the
trans-Planckian regime.
Finally, in Ref.~\cite{CGUP}, it was shown that the uncertainty in the
horizon size obtained from the HWF, combined with the standard
Heisenberg uncertainty for the particle's size, yields a minimum
detectable length which is always larger than the Planck scale.
In other words, a minimum detectable length of the order of the Planck
scale naturally emerges within the HWF formalism, without the need to
assume it, and without a modification of canonical commutation rules.
\subsection{Superextremal geometry}
We will now focus on studying overcharged sources,
represented by the range of specific charge $\alpha>1$.
It is well known that in the classical theory of gravity, the CCC
{\em a priori\/} forbids the existence of NSs.
In the case of the classical charged BHs, this would precisely correspond
to $\alpha > 1$, so it is interesting to investigate whether quantum physics
leads to any modifications or {\em predictions\/} therein.
Our guiding principle will be to assume that the quantum
states in the regime $\alpha>1$ can be obtained by extending continuously
the HWF from the case $\alpha<1$.
Of course, this is by no means the only possible choice, but one can 
at least hope that it leads to consistent predictions for $\alpha$ not
too much larger than the classical limiting value of $1$.
\par
The first problem that we encounter for $\alpha > 1$
is that the operators $\hat{R}_{\pm}$ are not Hermitian.
We could stop right there and say that there are no observables
which correspond to $\hat{R}_{\pm}$ in the regime $\alpha > 1$.
However, we will make a simple (and perhaps non-unique) choice,
which will allow us to proceed and probe the classically forbidden region. 
We will take the real parts of the multiplicative operators $\hat{R}_{\pm}$,
which are certainly Hermitian, to correspond to quantum observables. 
With this choice,  the modulus squared of the two HWFs
are given from Eq.~\eqref{psih}, for $R_\pm>R_{\rm min \pm}$, by
\be
\!\!\!\!\!\!|\psi_{\rm H}(R_{\pm})|^2
=
\mathcal{N_\pm}^2
\exp\left\{-\frac{\mpl^2\,R_{\pm}^2}{\Delta^2\,\lp^2\,(1\pm\sqrt{1-\alpha^2})^2}\right\},
\label{psih2}
\ee
which, for $\alpha>1$, becomes one expression 
\be
|\psi_{\rm H}(R)|^2
=
\mathcal{N}^2\,
\exp\left\{ - \frac{2-\alpha^2}{\alpha^4}\,\frac{\mpl^2 R^2}{\Delta^2 \lp^2}\right\}
\ ,
\label{psih1}
\ee
where $R$ now replaces both $R_+$ and $R_-$.
This HWF is still normalizable in the scalar product~\eqref{normH} if $R$
belongs to the real axis and the specific charge lies in the range
\be
1 < \alpha^2 < 2
\ .
\label{a>1}
\ee
We could therefore infer that no normalizable quantum state with $\alpha^2>2$
is allowed, or that there is an obstruction that prevents the system from
crossing $\alpha^2=2$.
This point will be further clarified after we have fully determined the HWF.
\par
The fact that the HWF~\eqref{psih1} is the same for $R_{+}$ and $R_{-}$ reflects 
the classical behaviour according to which the two real horizons merge at
the critical value of $\alpha=1$, and then mathematically extend as one into
the complex realm for $\alpha>1$.
We then need to address what happens to the Heaviside function in 
Eq.~\eqref{psih} when we extend it into the superextremal regime. 
First of all, we note that although Eq.~\eqref{Rminalpha} becomes complex
for $\alpha>1$, its real part is again the same for $R_+$ and $R_-$, namely
\be
R_{\rm min}
&=&
{\rm Re}\left[\lp\, \frac{m}{\mpl}\left(1\pm\sqrt{1- \alpha^2}\right)\right]
\nonumber
\\
&=&
\lp\, \frac{m}{\mpl}
\ .
\label{Rmin>1}
\ee
We can then show that the same continuity principle, which led us to
Eq.~\eqref{psih1}, requires that $R$ be bounded from below 
by this $R_{\rm min}$. 
In fact, the expectation value for $\hat R$ is, in this case,  
\be
&&
\expec{\hat R}
=
4\,\pi\,\int_{R_{\rm min}}^{\infty}
|\psi_{\rm H}(R)|^2\,R^3\,\d R
\label{expecRH}
\\
&&
=
\frac{2\,\lp^2\left(2-\alpha^2+\alpha^4\right)/\ell}
{\sqrt{2-\alpha^2}\,\left[2\,\sqrt{2-\alpha^2}+\alpha^2\,e^{\frac{2-\alpha^2}{\alpha^4}}\,\sqrt{\pi}
\,{\rm erfc}\left({\frac{\sqrt{2-\alpha^2}}{\alpha^2}}\right)\right]} 
\ .
\nonumber
\ee
and it matches exactly the corresponding expressions for $\alpha<1$~\cite{InProg},
\be
\expec{\hat R_{\pm}}
&=&
4\,\pi\,\int_{R_{\rm min\pm}}^{\infty}
|\psi_{\rm H}(R_{\pm})|^2\,R_{\pm}^3\,\d R_{\pm}
\nonumber
\\
&=&
\frac{4\,\lp^2\left(1\,\pm \sqrt{1-\alpha^2}\right)/\ell}{2\,+e\,\sqrt{\pi}\,{\rm erfc}{(1)}}
\ ,
\ee
namely
\be
\lim_{\alpha\searrow 1} \expec{ \hat R}
=
\frac{4\,\lp^2/\ell}{2+e\sqrt{\pi}\,{\rm erfc}{(1)}}
=
\lim_{\alpha\nearrow 1} \expec{ \hat R_{\pm}}
\ .
\label{alphato1}
\ee
One can likewise show that the uncertainty 
\be
\Delta R^2(\ell,\alpha>1)
=
\expec{\hat R^2} - \expec{\hat R}^2
\ ,
\label{DRH}
\ee
matches the corresponding uncertainties 
\be
\Delta R_\pm^2(\ell,\alpha<1)
=
\expec{\hat R_\pm^2} - \expec{\hat R_\pm}^2
\ ,
\label{DRHpm}
\ee
at the specific charge $\alpha=1$, but we omit the explicit expressions
since they are rather cumbersome.
We just note that, for $\alpha=1$, the width of the Gaussian
$\ell> \expec{\hat R}$ for 
$
m
<
{\sqrt{2+e\sqrt{\pi}\,{\rm erfc}{(1)}}}\,{\mpl}/2
\simeq
0.8\,\mpl
$,
so that quantum fluctuations in the source's size will dominate
for masses significantly smaller than the Planck scale (in qualitative
agreement with the neutral case~\cite{Casadio,C14,InProg}).
\par
\begin{figure}[t]
\centering
\includegraphics[width=8cm]{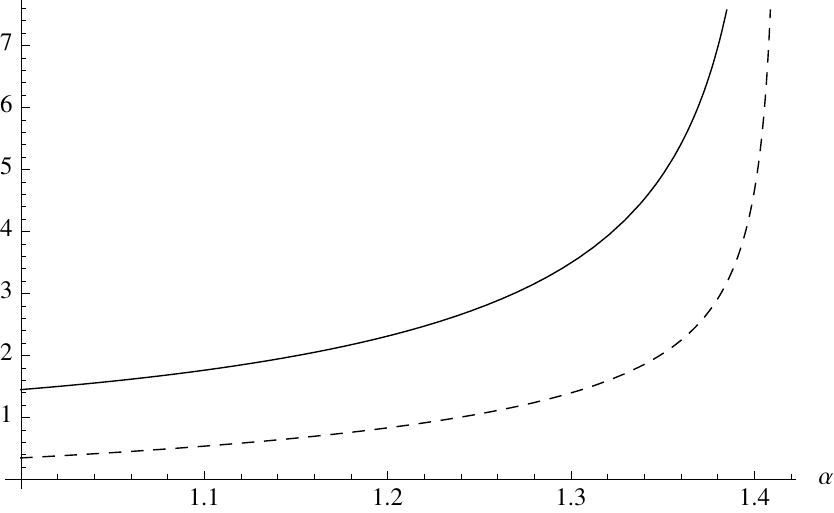}
\caption{The expectation value $\expec{\hat R}$ (solid line) and its uncertainty
$\Delta R$ (dashed line) as functions of the specific charge for $1<\alpha^2<2$ and
$m=\mpl$ ($\ell = \lp$).
\label{ExpR}}
\end{figure}

It is now interesting to analyse the limit $\alpha^2\to 2$.
One may have already noticed that 
\be
\expec{\hat R}
\simeq
\frac{2^{5/4}\,\lp^2/\ell}{\sqrt{\pi\,\left(\sqrt{2}-\alpha\right)}}
\ ,
\ee
so that the ratio $\expec{\hat R}/\ell$ blows up at $\alpha^2=2$
for any values of the mass $m=\mpl\,\lp/\ell$.
The same indeed occurs to the uncertainty, since 
\be
\Delta R
\simeq
\sqrt{{3\,\pi}/{8}-1}\,
\expec{\hat R}
\simeq
0.4\,
\expec{\hat R}
\ ,
\ee
for $\alpha^2\to 2$  (see also Fig.~\ref{ExpR}).
\par 
\begin{figure}[h]
\centering
\includegraphics[width=8cm]{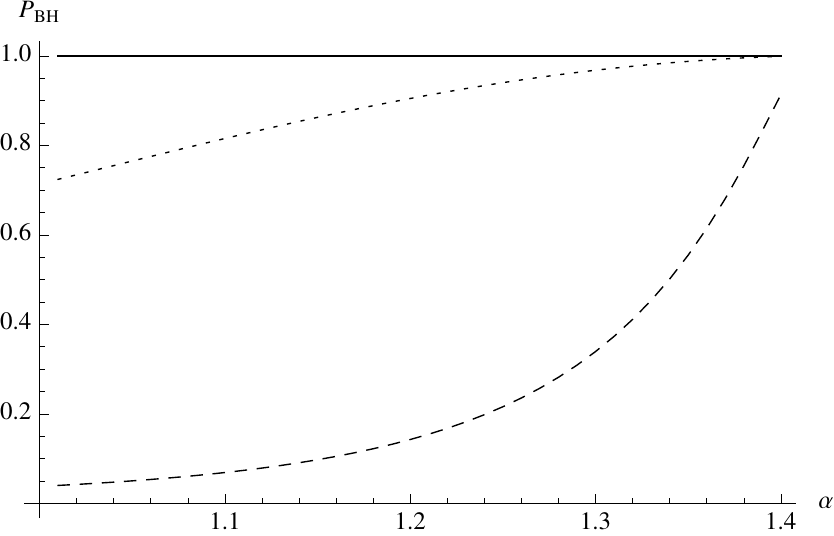}
\caption{$P_{\rm BH}$ as a function of $\alpha$ for $m = 2\, \mpl$
(solid line), $m= \mpl$ (dotted line) and $m=0.5\,\mpl$ (dashed line).
Cases with $m \gg  \mpl$ are not plotted since they behave
the same as  $m = 2\, \mpl$, i.e.~an object with $1<\alpha^2<2$
must be a BH.
\label{large alpha}}
\end{figure}
Using Eq.~\eqref{PBH+}, one can also calculate the probability $P_{\rm BH}$
that the particle is a BH for $\alpha$ in the allowed superextremal range~\eqref{a>1}.
This probability is displayed in Fig.~\ref{large alpha}.
One notices that, for a particle mass above the Planck scale, $P_{\rm BH}$
is practically one throughout the entire range of $\alpha$ (thus extending 
a similar result that holds for $\alpha<1$~\cite{InProg}).
Moreover, even for $m$ significantly less than  $\mpl$, $P_{\rm BH}$
approaches one in the limit $\alpha^2 \rightarrow 2$. 
We recall here that $P_{\rm BH}\ll 1$ for small $m$ is essentially related to
$\ell\gg \expec{\hat R}$, and the system is thus dominated by quantum fluctuations
in the source's position well below the Planck scale. 
On the other end, since both $\expec{\hat R}$ and $\Delta R$ blow up on
approaching $\alpha^2=2$, the superextremal configurations with a
significant probability of being BHs contain strong quantum fluctuations in
the horizon's size.
\par
Let us conclude this section by emphasising that, in order to
achieve the above results, we had to choose a way to continue the HWF
and operators that are straightforwardly defined for $\alpha<1$ into the
overcharged regime.
Since this choice is not apparently unique, one might wonder
whether different options would lead to significantly different outcomes.
Although we have not performed a full survey, it is important to remark
that we were able to ensure the continuity of expectation values
across $\alpha=1$, like in Eq.~\eqref{alphato1}, by imposing simultaneously
that $\hat R$ equals the real part of
$\hat R_\pm$ and Eqs.~\eqref{Rmin>1} and \eqref{psih1},
whereas we found no ways to make other smilingly natural options,
like $R=|R_\pm|$, work.
\section{Conclusions}
From the above analysis we can learn two important things.
First, quantum mechanical effects are perhaps able to continuously
take us into the classically forbidden region of $\alpha >1$.
This means that even an overcharged object, with a charge-to-mass
ratio greater than unity, can still make a quantum BH.
The basic reason for this is that in our formalism the location of
the horizon is not given by a sharp classical value, instead it is
described by a quantum wave function with associated uncertainties.
Second, the charge-to-mass ratio $\alpha$ cannot be arbitrarily
large, even in the context of QM.
We found that for $\alpha^2 > 2$ the HWF cannot be normalised,
and thus it is not describing a well defined physical object.
Moreover, at the same value of $\alpha^2=2$, the uncertainties in the
location of the horizon become infinite, signalling again that
such an object stops being well defined.
We should warn the reader that the specific value of the upper limit
$\alpha^2=2$ in Eq.~\eqref{a>1} might simply be a consequence 
of describing the source as the Gaussian function~\eqref{psis},
and should not be taken literally.
However, it is likely that the overall qualitative picture remains 
in a more general context, and our results imply that perhaps a
{\em quantum\/} version of the CCC might be formulated by stating
that no BHs with the charge-to-mass ratio greater than a critical
value (of order $\sqrt{2}$) can exist.
\par
We should here recall that for the charged Reissner-Nordstr\"om
metric with $\alpha\le 1$ analysed in Ref.~\cite{InProg}, as well for
neutral sources~\cite{C14,CGUP,BEC_BH}, the single Gaussian
constituent~\eqref{psis} leads to unacceptably large uncertainties
in the horizon size of large astrophysical BHs.
In fact, one has $\Delta R\sim \expec{\hat R}$, even for very 
large mass $m$, for which one expects a semiclassical behaviour for
the horizon size $R$.
The above quantum CCC will therefore have to be tested further,
by considering models of BHs that allow for a semiclassical limit
$\Delta R\ll \expec{\hat R}$.
An example of such models is given by those in Refs.~\cite{dvali,BEC_BH}, 
which contain a very large number $N$ of light constituents, whose wave-functions
span the entire region inside $R$, and $\Delta R/\expec{\hat R}\sim N^{-1}$.
The emerging picture is that BHs of any size should be treated
as macroscopic quantum objects (just like superconductivity and superfluidity
are macroscopic quantum phenomena at scales where one expects classical
physics to be a good description).
\par
Finally, let us point out that the analysis performed in this work, and the
above quantum CCC, should hold for sources with mass $m$
within a few orders of magnitude of the Planck mass. 
Primordial BHs formed in the early universe by large density
fluctuations could have masses in this range.
Moreover, it is also plausible that overcharged configurations with such small
masses emerge from the gravitational collapse of astrophysical objects,
acting as seeds for much larger BHs.
Our results should then apply straightforwardly to these two cases.
\subsection*{Acknowledgments}
R.C.~was supported in part by the INFN grant FLAG.
O.M.~was supported in part by research grant UEFISCDI
project PN-II-RU-TE-2011-3-0184.
D.S.~was partially supported by the US National Science Foundation,
under Grant Nos.~PHY-1066278 and PHY-1417317.

\begin{thebibliography}{99}
%
%
\bibitem{OS}
J.R.~Oppenheimer and H.~Snyder,
Phys.\ Rev.\  {\bf 56} (1939) 455;
J.R.~Oppenheimer and G.M.~Volkoff,
Phys.\ Rev.\  {\bf 55}  (1939)  374.
%
\bibitem{joshi}
P.S.~Joshi,
``Gravitational Collapse and Spacetime Singularities,''
Cambridge Monogr.~Math.~Phys.,
Cambridge University Press (Cambridge, 2007).
%
\bibitem{Penrose:1969pc}
  R.~Penrose,
  Riv.\ Nuovo Cim.\  {\bf 1} (1969) 252
   [Gen.\ Rel.\ Grav.\  {\bf 34} (2002) 1141].
  %
  %
\bibitem{Greenwood:2008ht}
  E.~Greenwood and D.~Stojkovic,
 JHEP {\bf 06} (2008) 042;
%
  A.~Saini and D.~Stojkovic,
  Phys.\ Rev.\ D {\bf 89} (2014) 044003;
%
  T.~Vachaspati, D.~Stojkovic and L.~M.~Krauss,
  Phys.\ Rev.\ D {\bf 76} (2007) 024005;
%
  T.~Vachaspati and D.~Stojkovic,
  Phys.\ Lett.\ B {\bf 663}  (2008) 107.
%
\bibitem{Wang:2009ay} 
  J.~E.~Wang, E.~Greenwood and D.~Stojkovic,
  Phys.\ Rev.\ D {\bf 80} (2009) 124027.
%
\bibitem{Casadio} 
R.~Casadio,
``Localised particles and fuzzy horizons: A tool for probing Quantum Black Holes,''
arXiv:1305.3195 [gr-qc];
%
\bibitem{C14}
R.~Casadio,
Eur.\ Phys.\ J.\ C {\bf 75} (2015) 160.
%
\bibitem{CGUP} 
R.~Casadio and F.~Scardigli,
Eur.\ Phys.\ J.\ C {\bf 74}  (2014) 2685.
%
\bibitem{Ctest}
R.~Casadio, O.~Micu and F.~Scardigli,
Phys.\ Lett.\ B {\bf 732} (2014) 105.
%
\bibitem{BEC_BH}
R.~Casadio, A.~Giugno, O.~Micu and A.~Orlandi,
Phys.\ Rev.\ D {\bf 90} (2014) 084040.
%
\bibitem{InProg}  
R.~Casadio, O.~Micu and D.~Stojkovic,
``Inner horizon of the quantum Reissner-Nordstr\"om black hole'',
to appear in JHEP, 
arXiv:1503.01888 [gr-qc].
%
%
\bibitem{dvali}
G.~Dvali and C.~Gomez,
``Black Holes as Critical Point of Quantum Phase Transition'',
arXiv:1207.4059 [hep-th];
Phys.\ Lett.\ B {\bf 719}  (2013) 419;
Phys.\ Lett.\ B {\bf 716}  (2012) 240;
Fortsch.\ Phys.\  {\bf 61}  (2013)742;
%
R.~Casadio and A.~Orlandi,
JHEP {\bf 1308}  (2013) 025.
%
\end{thebibliography}
\end{document}